\documentclass[multphys,vecphys]{svmult}

\usepackage{makeidx}         % allows index generation
\usepackage{graphicx}        % standard LaTeX graphics tool
\usepackage{multicol}        % used for the two-column index
\usepackage[bottom]{footmisc}% places footnotes at page bottom
\makeindex             % used for the subject index
\begin{document}

\title*{Radio Galaxies in Cooling Cores:  Insights from a Complete Sample}
% Use \titlerunning{Radio Galaxies in Cooling Cores} for an abbreviated 
% version of
\author{Jean A. Eilek\inst{1}\and
Frazer N. Owen\inst{2}}
\institute{New Mexico Tech, Socorro NM USA \texttt{jeilek@aoc.nrao.edu}
\and NRAO, Socorro NM USA \texttt{fowen@aoc.nrao.edu}}
\maketitle

\begin{abstract}

We have observed a new, complete, cooling-core sample with the VLA, in order
to understand how the massive black hole in the central galaxy interacts
with the local cluster plasma.  We find that every cooling core is 
currently being energized by an active radio jet, which has probably
been destabilized by its interaction with the cooling core.   
We argue that current  models of cooling-core radio galaxies
need to be improved before they can be used to determine the rate at which
the jet is heating the cooling core.   We also argue that the extended radio
haloes we see in many cooling-core clusters need extended, {\it in situ}
re-energization, which cannot be supplied solely by the central 
galaxy.

\end{abstract}

What heats cooling cores?  The spotlight has turned on active galactic
nuclei (AGN),  driven by massive black holes in the heart of the 
 galaxy at the center of the cooling core (CC).  The jets in a few
 bright, well-studied radio galaxies (RGs) ({\it e.g.}, M87 
\cite{OEK}; A2052 \cite{Blanton};  Perseus A  \cite{Fabian})  seem to be
pouring out more than enough energy to offset radiative   cooling in
the CC in these clusters.  But is this not the full answer;  questions
remain.   Does every CC have a central RG? Is a typical cooling-core
radio galaxy (CCRG) strong enough to offset local cooling?  Does the energy
carried by the jet couple effectively to the intracluster medium (ICM)?  
 How can
we use radio and X-ray data to estimate the jet power and energy
input to the CC? To answer
these questions, we must  study more than the brightest few CCRGs, 
and must also look critically at dynamical models of the RG.  We therefore
carried out deep radio observations of a complete  sample of CCRGs.
In this paper we summarize our results and speculate on how to extend
current models;  more details  will be given in \cite{EOM}.

\section{The Data: What We Did}

We formed a complete, X-ray selected sample of CCs in
nearby Abell clusters.  We started with ROSAT All-Sky Survey
images of nearby ($z < 0.09$) Abell clusters \cite{Ledlow}.  We
identified clusters which are X-ray bright ($L_x > 3 \times 10^{43}$erg/s
within a 500 kpc aperture), centered on a massive galaxy, and with a centrally
concentrated X-ray atmosphere (ratio of flux within 500 and 62.5 kpc
apertures no larger than $\sim 14$).  These criteria correlate well
with strong CCs found in other, more detailed deprojection analyses ({\it
  e.g.}, \cite{Peres}.)  From these we selected clusters favorably
placed in the sky for nightime VLA observations in 2002. 
This procedure gave us 22 clusters:  A85\index{A85}, A133\index{A133},
A193\index{A193}, A426\index{A426}, A496\index{A496},
A780\index{A780}, A1644\index{A1644}, A1650\index{A1650}, A1651\index{A1651},
A1668\index{A1668}, A1795\index{A1795}, A1927\index{A1927},
A2029\index{A2029}, A2052\index{A2052}, A2063\index{A2063}, 
A2142\index{A2142}, A2199\index{A2199}, A2428\index{A2428},
A2495\index{A2495},  A2597\index{A2597},
A2626\index{A2626} and A2670\index{A2670}.   
  Good, deep radio data already exist for 3 of these
(A85;  A426, the Perseus A cluster;  A780, the Hydra A cluster).  
We observed the
rest with the VLA.  Because high-resolution radio data exist for
many of these objects, we designed our observations to detect faint,
extended radio emission \cite{Tomislav}.  We note that  M87\index{M87} 
is not in our  formal sample, because it is not in an
Abell cluster, and its CC is on the weak side.  We include it in much
of our analysis, however, because it is  so well studied \cite{OEK,
Forman}, and it is  an important example of the interaction
between CCs and their embedded RGs.   In addition, nine of our 
clusters were also 
included in our VLA search \cite{Tomislav} for  cluster-scale
radio haloes, giving us additional information on extended emission
from these objects. 

\section{The Data: what we have learned}

Our data show that the story is more complicated than has been thought.
We find evidence every that cooling core is being disrupted, and probably
energized, by an AGN.  Our data also suggest that the radio-loud
plasma does mix with the ICM, at least on large scales, and that the
AGN may well {\it not} be the only driver for the ICM in a cooling-core
 cluster.

\subsection{Every Cooling Core Contains a Radio Source}

Every cooling core in our sample contains a currently-active
radio core (some too faint to have been detected in previous
work). This means that the central AGN are active  100\%
of the time; they do not have any ``off'' periods.
However, if currrent dynamical estimates of source ages ($\sim 100
$ Myr) are close to correct, the central AGN is probably variable,
cycling through high-power and low-power states.  The    
jet and inner halo of M87\index{M87} \cite{Dean} appear to be an
example of a recently  ``reinvigorated'' AGN.   

It follows that  every RG in a cooling-core is currently being
driven by an active jet.  In particular, our deep radio images
sometimes reveal  faint jets connecting the central AGN to what were
previously thought to be offset ``relics'' (e.g., A133\index{A133},
A2199\index{A2199}).  We therefore argue that very few CCRGs
 are simply passive, buoyant bubbles.  The situation is more complex;  
we hope the data can guide us toward improved models. 

\subsection{The Radio Galaxies are Unusually Disturbed}

Cooling-core RGs are characterized by unusual
morphologies.  Most of them are neither Fanaroff-Riley Type I
(tailed), nor Type II (classical doubles).  Sixteen CCRGs in
our sample are well enough imaged to reveal their structure (the remaining 
7 are too faint and too small).   Three of these 16 are
standard tailed sources (including Hydra A).  The rest are
diffuse and amorphous. Such a large fraction (80\% of the set) is far 
too many to be  ``normal'' RGs seen end-on. 
Furthermore,  amorphous sources such as these are rare in the general 
radio-galaxy population;  only 5 of the $\sim 200$ well-imaged cluster
radio galaxies in the Owen-Ledlow set \cite{OL} are amorphous, and
{\it all} of those sit in the centers of strong cooling cores.   

These data show that an unusually strong interaction occurs between
the radio jet and the dense cooling-core into which it tries to
propagate.  The interaction seems to destabilize the jet, on a scale
of only a few kpc (the short jets in M87\index{M87}, Perseus A\index{PerA}
and A2052\index{A2052}  are good examples here).
It follows that the evolution of a cluster-core RG is not 
governed by directed momentum flux, in a large-scale jet, as is the
case with most RGs.  Instead,  isotropized energy flow from a
disrupted jet creates the amorphous haloes that we see.  The strong
RG-CC interaction is also suggestive of an effective energy transfer
between the jet and the local cooling core;  however 
the details of the process remain unclear.

\subsection{The Radio Haloes Extend to Large Scales}

Many of our  CCRG are much larger than was previously known.  
Most  of the amorphous sources have two
scales of radio emission: a  smaller, brighter source (often 
previously studied in higher resolution observations)  
is embedded within a larger, faint, extended mini-halo. 
Typical sizes of these mini-haloes 
range from $\sim 70$ to $ 200$ kpc. For instance, Per A \cite{Burns} and 
A2029 \cite{Tomislav} can be traced to $\sim 200$ kpc, nearly as large
as the long radio tails of Hyd A \cite{Lane}. 
If the AGN is always ``on'', but cycles between strong and weak states, the
mini-haloes  may be  relics of previous activity cycles.
In addition, we have detected 
Mpc-scale radio haloes in two clusters, A2328 and A2495, which
refutes the  current idea that Mpc-scale haloes avoid 
cooling cores.

To put these sizes in context,  recall that the  size of the cooling
core is typically $\sim50\!-\!100$ kpc.  The size 
of the cluster's potential well, as measured by the Navarro-Frenk-White
scale radius,  is only a few hundred kpc in CC clusters. 
Furthermore, the haloes do not obviously have clear edges. 
The sizes we measure are limited by the sensitivity of the observations,
and may not be the true extent of the radio emission. 
With such large scales, the radio haloes may better be regarded as part of
the entire ICM, not just a byproduct of the central AGN.

\subsection{The Radio and X-ray Plasmas Must Mix}

Our sample contains a variety of mixing states. Some of our
clusters have small, clear  X-ray cavities in the inner CC, approximately 
coincident with the RG.  These cavities are very likely filled by
radio-loud plasma, with little or no X-ray plasma.  In other clusters
({\it e.g.},  M87 \cite{Forman},  A1795 \cite{Ettori}),
the interaction between the  two plasmas is more complex. The X-ray
plasma is clearly interacting with, but not evacuated  by, the radio
plasma.  In still others, the X-ray
plasma appears smooth and undisturbed on the scale of the RG, at the
best current  X-ray resolution and  
sensitivity.   We suspect the two plasmas have at least partially mixed in
these CCs.  In addition, because the ICM is not
dramatically disturbed on  hundred-kpc scales, the larger radio
haloes must be effectively mixed with the ICM.

It seems, therefore, that the radio and X-ray plasmas mix effectively during
the lifetime of the CCRG.   Just how
this occurs is not clear,  given the stabilizing effects of even a small 
magnetic field in the ICM \cite{JdeY}.  
Deep, high resolution
radio images ({\it e.g.}, M87, \cite{OEK};  A2199, in preparation) may 
provide hints.  We see radio-loud filaments in these sources 
that appear to be escaping from the RG and penetrating the CC 
plasma.  These filaments may be
similar to the magnetic flux  ropes which are
known to penetrate the terrestrial  magnetopause. % (e.g., \cite{Pu}). 
Such flux ropes, once formed, may 
 decouple from the main body of the radio source and rise bouyantly 
through the ICM, giving rise to an extended radio halo coincident with
a relatively smooth X-ray atmosphere.

\section{The Models:  Where to Go Next}

The important question in the context of this meeting is, how much
energy does a ``typical'' central AGN  deposit in the
cooling core plasma? 
To answer this, we must  determine the mean jet power,
$P_j$, averaged over the lifetime of a typical CCRG, and how
effectively that  power is deposited in the local plasma.   
We emphasize that we have {\it no} direct measure of $P_j$.  The radio
power of the CCRG is a poor tracer of the jet power \cite{JAE}.
 The best we can do directly from observations is to use minimum-pressure
arguments, which are possible if the jet is  resolved ({\it e.g.} M87
\cite{OEK}).  This gives us a lower bound on $P_j$.  To go further,
we must choose a dynamical model for the RG and its interaction with the
local ICM.  This sounds simple, but the devil is in the details.

\subsection{Calorimetry}

The simplest cases are cooling cores which have clear X-ray cavities 
that coincide  with an extended RG.  In these the
mean jet power can be found from the
energy within the cavity and the age of the source.  This is a simple, 
attractive appproach, which  has been applied by various authors 
({\it e.g.},  \cite{Birzan, DeY}).  However, it has complications.  One is that
measuring the energy  content of the cavity is not straightforward, 
because it is hard to know the extent to which the radio and X-ray plasmas 
have mixed in most of these
clusters.  A second concern is how to estimate the age of the source.  
Most authors currently assume the radio source is passive, having been
previously inflated by an AGN which has since turned off.  If
this holds, the source age is its size divided by the bouyant speed, $v_b$.
 But what is $v_b$? 
The sound speed is no more than an optimistic upper limit, because
$v_b$ is quite subsonic for small structures.  In  addition,  magnetic 
tension from even a very small intracluster 
field can  exceed hydrodynamic
drag, and reduce $v_b$ even more \cite{JdeY}.

A more serious concern, however, is the evidence from our work
 that every CCRG is currently being  driven by jets from an active AGN,
and  that large-scale radio haloes have mixed with the ICM. 
It follows that very few CCRGs are well described as isolated,
passive, buoyant bubbles 
(although buoyancy surely plays  some role in the evolution
of the RG).  New models are needed.

\subsection{Possible Dynamical Models}  

As a first step towards such new models, we
suggest  that CCRG evolution can be broken into two stages.  
We envision an early stage in which young, driven sources interact 
with and expand into the ICM, 
and a later stage in which the RGs have grown into extended mini-haloes
mixed with the ICM.  If AGN activity is cyclic,
an older mini-halo could coexist with a younger, restarted inner core.
We note that our ideas here are no more than toy models; they need to be
developed and tested  against real, well-observed CCRGs.

Because the observations show that the AGN in every CCRG is ``alive'', and
because CCRGs are often
amorphous, we suggest that small, young sources
are  being driven by a quasi-isotropic energy flux (as from
an unstable jet). Such evolution can be approximated by a self-similar analysis 
\cite{Falle}.  However, because the edges of the 
X-ray cavities are not strong shocks ({\it e.g.},  \cite{Fabian}), we know
the expansion is slow;  this suggests the expansion proceeds at approxiate
pressure balance \cite{OEK}.  Such a  model predicts the source size 
$R(t)  \propto (P_j t)^x$, where $x$ depends on the ambient pressure
gradient.  We emphasize that   $P_j$ and $t$ cannot
be determined separately in this model;  the 
best we can do is the limit $\dot R < c_s$, which gives an upper limit
to $P_j$. 

Because the data also show  the radio and X-ray plasmas are well  mixed 
for larger RGs, we further suggest that CCRGs eventually fragment and mix
 with the ICM.    The fragmentation may occur {\it via}  MHD 
surface effects (such as the tearing-mode instability) which create magnetic 
filaments or flux ropes.  Alternatively,  the small-scale flux ropes which
we know exist in MHD turbulence may  retain coherence and diffuse
into the extended ICM in late stages of CCRG evolution.  (The 
ubiquity of filaments in well-imaged RGs suggests such structures
are common in general; why should CCRGs be different?)
We expect the flux ropes to rise slowly under buoyancy, and to retain their
identity for awhile, after which they probably dissipate and merge with
the local ICM. In principle, $P_j$ could be estimated for such a 
source from the energy content of the radio plasma and its
buoyant rise time, but uncertainties in filling factors and flux 
rope sizes limit the quantitative usefulness of this  approach.

\subsection{The Large Radio Haloes}  

Some of our radio haloes are large enough to raise the question, 
where does ``CCRG'' end and ``cluster halo'' begin?  That is, on what
scale is the physics of the  full cluster more important than the 
influence of the AGN?   The synchrotron size is one criterion:  how large 
can the radio halo can be without needing {\it in situ} energization? 
We are skeptical of simple 
synchrotron-aging estimates, because magnetic fields in the radio source and
the ICM are  almost certainly  inhomogeneous.   One can, however, derive
a useful limit.  The lowest loss rate for the radio-loud electrons is that of 
inverse Compton losses on the cosmic microwave background.  
If the electrons spend  most of their time
 in sub-$\mu$G magnetic  fields, and occasionally migrate into 
high-field regions (probably a few $\mu$G) 
where they become radio-loud, we can find  an upper limit to 
their synchrotron life. 

This cartoon predicts the radio plasma in a buoyant 
flux rope can reach $\sim 100$ kpc before it fades away.
 Radio sources larger than this  must be undergoing
extended, {\it in situ} re-energization.  It follows that some driver other
than the AGN must exist on large scales.  
Ongoing minor mergers are thought to support
radio haloes in large, non-CC clusters.  They may be important 
in CC clusters as well  ({\it e.g.}, \cite{Motl}), and may be driving the 
larger haloes.  But then, if we admit the need
for non-AGN heating of CC clusters on large scales, can we be  sure that the
cooling core itself is heated only by the AGN?

\printindex
\end{document}